\documentclass{PoS}

\title{The two-nucleon problem in EFT reformulated: \\ Pion and nucleon
  masses as soft and hard scales}

\ShortTitle{The NN problem in EFT reformulated}

\author{E.~Epelbaum
\\
         Institut f\" ur Theoretische Physik II, Fakult\" at f\" ur Physik und Astronomie,\\ Ruhr-Universit\" at Bochum 44780 Bochum, Germany\\
        E-mail: \email{Evgeny.Epelbaum@tp2.ruhr-uni-bochum.de}}

\author{\speaker{J.~Gegelia}\\
        Institut f\" ur Theoretische Physik II, Fakult\" at f\" ur Physik und Astronomie,\\ Ruhr-Universit\" at Bochum 44780 Bochum, Germany\\  Tbilisi State University, 0186 Tbilisi, Georgia\\
        E-mail: \email{Jambul.Gegelia@tp2.ruhr-uni-bochum.de}}

\abstract{We outline the modified formulation of baryon chiral
  effective field theory for nucleon-nucleon scattering and discuss
  the
issue of a possible power counting violation by the nucleon mass.
We also present the results for the quark mass dependence of the $^1S_0$ and $^3S_1$
scattering lengths and the  deuteron binding energy at leading order.
}

\FullConference{The 7th International Workshop on Chiral Dynamics,\\
		August 6 -10, 2012\\
		Jefferson Lab, Newport News, Virginia, USA}

\begin{document}
\section{The method}

A systematic quantum field theoretical approach to few-nucleon systems
has been pioneered by Weinberg in Ref.~\cite{Weinberg:rz}, (see
Refs.~\cite{Epelbaum:2008ga,Machleidt:2011zz} for recent review articles). Within this
framework, the  nucleon-nucleon (NN) potential is defined as a sum of
two-nucleon-irreducible time ordered diagrams emerging in
non-relativistic effective field theory (EFT). It is calculated as a
series based on  a systematic expansion in small parameters.
A finite number of diagrams contribute to the effective potential at any finite order.
The scattering amplitude is obtained by solving the Lippmann-Schwinger
(LS) or, equivalently, Schr\"odinger equation.

The problem of renormalization turned out to be highly non-trivial  in
Weinberg's approach. To resolve this problem we have recently suggested
a new framework based on the manifestly Lorentz invariant effective
Lagrangian and time ordered perturbation theory
\cite{Epelbaum:2012ua,Epelbaum:2012cv}.
In this approach the leading-order amplitude is obtained by solving
the integral equation
\begin{equation}
{T _{0 }\left(
\vec p\,',\vec p\right)}{=}{ V _{0 }\left(
\vec p\,',\vec p\right) - \frac{m^2}{2}\, \int \frac{d^3 k}{(2\,\pi)^3}} {\frac{V _{0 }\left(
\vec p\,',\vec k\right) \, T _{0 }\left(
\vec k,\vec p\right)}{\left(k^2+m^2\right)\, \left(
E-\sqrt{k^2+m^2}+i\,\epsilon\right)},}
\label{kad}
\end{equation}
where $E=\sqrt{p^2+m^2}$ denotes the energy of a single nucleon in the
center of mass frame. Here and in what follows, we use the
notation $p \equiv | \vec p \, |$, $k \equiv | \vec k \, |$.
The leading order (LO) NN potential can be taken in the usual form
\begin{equation}
V_0\left(
\vec p\,',\vec p\right) = C_S+C_T\, \vec\sigma_1\cdot\vec\sigma_2
-\frac{g_A^2}{4\,F^2}\ \vec\tau_1\cdot\vec\tau_2 \
\frac{\vec\sigma_1\cdot\left(
\vec p\,'-\vec p\right)\,\vec\sigma_2\cdot \left(
\vec p\,'-\vec p\right)}{\left(
\vec p\,'-\vec p\right)^2+M_\pi^2},
\label{LOV}
\end{equation}
where the standard notation is employed, see \cite{Epelbaum:2012ua}
for more detail{\bf s}. Notice that equation (\ref{kad}) was first obtained
in Ref.~\cite{kadyshevsky}.
Its iterations for the potential (\ref{LOV}) generate only logarithmic
divergences which can be absorbed into redefinition of the couplings
$C_S$ and $C_T$, i.e. it is perturbatively renormalizable. Partial
wave projected equations corresponding to Eq.~(\ref{kad}) have unique
solutions except for the $^3P_0$ channel.  We solved the problem of
non-uniqueness of the solution in this partial wave analogously to
Ref.~\cite{Bedaque:1998kg} by including a counter term of the form
$C(\Lambda) p' p/\Lambda^2$, with $C(\Lambda)$ being a cutoff dependent
constant,  in the leading-order potential \cite{Epelbaum:2012ua}.

\section{Iterations of one-pion exchange and the role of the
  nucleon mass}

Comparing Eq.~(\ref{kad}) with the Lippmann-Schwinger equation of the
non-relativistic heavy baryon
chiral perturbation theory \cite{Weinberg:rz} one might come to the
(wrong) conclusion that the nucleon
mass plays the role of the cutoff in Eq.~(\ref{kad}) and, being the
hard scale, violates the power counting.
We now analyze the nucleon-mass dependence of
renormalized loop diagrams by considering the first two iterations of
the one-pion exchange (OPE)
potential in order to demonstrate explicitly that the above naive
interpretation is misleading.

We first
consider the dimensionally regularized
one-loop tensor integral corresponding to a single iteration of the OPE potential
\begin{equation}
I_1 =-\frac{m^2}{2}\int \frac{d^nk\ \left(p'_a-k_a\right)\left(p'_b-k_b\right)\left(p_i-k_i\right)\left(p_j-k_j\right)}{
   \left(k^2+m^2\right)
   \left(E-\sqrt{k^2+m^2} \right)\left[(p'-k)^2+M^2\right]
   \left[(p-k)^2+M_\pi^2\right]},
\label{1it}
\end{equation}
where we drop, for the sake of simplicity, the
$+i\,\epsilon$ prescription for two-nucleon propagators.
Its expansion in inverse powers of the nucleon mass can be easily obtained by applying the method of dimensional
counting \cite{Gegelia:1994zz} and has the form
\begin{eqnarray}
I_1 &=& m^2 m^{n-3}\int \frac{d^nq\ q_a q_b q_i q_j}{
  2  q^4 \left[q^2+1\right] \left[\sqrt{q^2+1}-1 \right]}+\ldots,
\label{1it}
\end{eqnarray}
where $q$ is a (re-scaled) dimensionless integration variable and the ellipses refer to terms
of higher orders in the $1/m$-expansion. The
first term in Eq.~(\ref{1it}) is momentum and pion-mass independent
and is cancelled  by a counter-term associated with the coupling
constants of the LO contact interactions. Higher-order terms in
Eq.~(\ref{1it}) have either the same linear dependence on the nucleon
mass (for spacetime dimension $n=3$) as in the heavy-baryon expression, or are suppressed by additional
inverse powers of $m$.  Thus, the nucleon mass does
not violate the power counting in our new approach at one loop level.

We now turn to the two-loop tensor integral emerging from the second iteration of the OPE potential
\begin{eqnarray}
I_2 &=&\frac{m^4}{4}\int \frac{d^nk_1 d^n k_2
  \left(p'_a-k_{1a}\right)\left(p'_b-k_{1b}\right)\left(k_{1i}-k_{2i}\right)\left(k_{1j}-k_{2j}\right)\left(k_{2\mu}-p_\mu\right)\left(k_{2\nu}-p_\nu\right)}{
   \left(k_1^2+m^2\right)
   \left(k_2^2+m^2\right)
   \left(E-\sqrt{k_1^2+m^2} \right)
   \left(E-\sqrt{k_2^2+m^2} \right) }\nonumber\\
   &\times& \frac{1}{\left[(p'-k_1)^2+M_\pi^2\right] \left[(k_1-k_2)^2+M_\pi^2\right]\left[(p-k_2)^2+M_\pi^2\right]}.
\label{int2input}
\end{eqnarray}
Its expansion in inverse powers of the nucleon mass has the form \cite{Gegelia:1994zz}
\begin{eqnarray}
I_2 & = &
\frac{m^4 m^{2 n-6}}{4}\int \frac{d^nq_1 d^n q_2\
  q_{1i}q_{1j}q_{2\mu}q_{2\nu}\left(q_{1a}-q_{2a}\right)\left(q_{1b}-q_{2b}\right)}{
 q_1^2\left(1+q_1^2\right)
 \left[\sqrt{1+q_1^2}-1\right]q_2^2\left(1+q_2^2\right)
 \left[\sqrt{1+q_2^2}-1\right](q_1-q_2)^2}\nonumber\\
&+&\frac{m^3 m^{n-3}}{2}\int \frac{d^nq_1 d^n k_2\
  q_{1a}q_{1b}q_{1i}q_{1j}\left(k_{2\mu}-p_\mu\right)\left(k_{2\nu}-p_\nu\right)}{
\left(k_2^2-p^2\right)q_1^4\left(1+q_1^2\right)
\left[\sqrt{1+q_1^2}-1\right]\left[(p-k_2)^2+M_\pi^2\right]}\nonumber\\
&+&\frac{m^3 m^{n-3}}{2}\int \frac{d^nq_2 d^n k_1\
  q_{2a}q_{2b}q_{2\mu}q_{2\nu}\left(k_{1i}-p'_i\right)\left(k_{1j}-p'_j\right)}{
\left(k_1^2-p^2\right)q_2^4\left(1+q_2^2\right)
\left[\sqrt{1+q_2^2}-1\right]\left[(p-k_1)^2+M_\pi^2\right]}+\ldots \,.
\label{int2Exp}
\end{eqnarray}
Renormalization of two-loop diagrams requires the addition of one-loop
diagrams generated by a single iteration of the one-loop counter
terms, see Eq.~(\ref{1it}), in order to
subtract the sub-divergences (and finite pieces) of one-loop
sub-diagrams. Correspondingly, we need to add  to the integral $I_2$ two
counter-term integrals:
\begin{eqnarray}
I_{2ct} & = & \frac{m^4 m^{n-3}}{4}\Biggl\{\int \frac{d^nq d^n k_1\
  q_{a}q_{b}q_{\mu}q_{\nu}\left(p'_i-k_{1i}\right)\left(p'_j-k_{1j}\right)}{
(k_1^2+m^2)\left(\sqrt{p^2+m^2}-\sqrt{k_1^2+m^2}\right)
q^4\left(1+q^2\right)
\left[\sqrt{1+q^2}-1\right]\left[(p'-k_1)^2+M_\pi^2\right]}\nonumber\\
&+& \int \frac{d^nq d^n k_2\ q_{a}q_{b}q_{i}q_{j}\left(p_\mu-k_{2\mu}\right)\left(p_\nu-k_{2\nu}\right)}{
(k_2^2+m^2)\left(\sqrt{p^2+m^2}-\sqrt{k_2^2+m^2}\right)
q^4\left(1+q^2\right)
\left[\sqrt{1+q^2}-1\right]\left[(p-k_2)^2+M_\pi^2\right]}\Biggr\}.
\label{ctds}
\end{eqnarray}
The expansion of these counter-term integrals in inverse powers of the
nucleon mass has the form
\begin{eqnarray}
I_{2ct}&=&
-\frac{m^4 m^{2 n-6}}{4} \Biggl\{ \int \frac{d^nq_2 d^n q\ q_{2\mu}q_{2\nu} q_{i} q_{j} q_{a} q_{b}}{
 q_2^2\left(1+q_2^2\right) \left(\sqrt{1+q_2^2}-1\right) q^4\left(1+q^2\right) \left(\sqrt{1+q^2}-1\right)}\nonumber\\
&+& \int \frac{d^nq_1 d^n q\ q_{1i} q_{1j} q_{\mu} q_{\nu} q_{a} q_{b}}{
 q_1^2\left(1+q_1^2\right) \left(\sqrt{1+q_1^2}-1\right) q^4 \left(1+q^2\right) \left(\sqrt{1+q^2}-1\right)}
 \Bigg\}
\nonumber\\
&-&\frac{m^3 m^{n-3}}{2}\Biggl\{ \int\frac{d^nq d^n k_2\ q_{a}q_{b}q_{i}q_{j}\left(k_{2\mu}-p_\mu\right)\left(k_{2\nu}-p_\nu\right)}{
\left(k_2^2-p^2\right)q^4\left(1+q^2\right) \left(\sqrt{1+q^2}-1\right)\left[(p-k_2)^2+M_\pi^2\right]}\nonumber\\
&+& \int \frac{d^nq_2 d^n k_1\ q_{a}q_{b}q_{\mu}q_{\nu}\left(k_{1i}-p'_i\right)\left(k_{1j}-p'_j\right)}{
\left(k_1^2-p^2\right)q^4\left(1+q^2\right) \left(\sqrt{1+q^2}-1\right)\left[(p-k_1)^2+M_\pi^2\right]}\Biggr\}+\ldots \,.
\label{ctdexp}
\end{eqnarray}
Adding the expressions in Eqs.~(\ref{int2Exp}) and (\ref{ctdexp})
together we observe that all terms proportional to $m^3 m^{n-3}$
cancel exactly. The remaining terms proportional to $m^4 m^{2 n-6}$
are momentum and pion-mass independent. They are subtracted by
the two-loop counter-terms generated by coupling constants of the LO
contact potential. The remaining terms have either the
same quadratic dependence on the nucleon mass as in the heavy-baryon
approach or are suppressed by additional inverse powers of $m$.
Analogously,  it can be shown
for any number of iterations
that the hard dependence on the nucleon mass is removed  by
renormalization yielding the amplitude which obeys the standard power
counting of the heavy-baryon approach. The naive interpretation of the
nucleon mass playing the role of the cutoff in our new formulation is
misleading because it is based on
the comparison with the heavy-baryon result ignoring the crucial
fact that the heavy-baryon expansion does not commute with the
expansion in inverse powers of the cutoff parameter.

\section{Pion-mass dependence at leading order}

\begin{figure*}
\begin{center}
\includegraphics[width=0.5\textwidth]{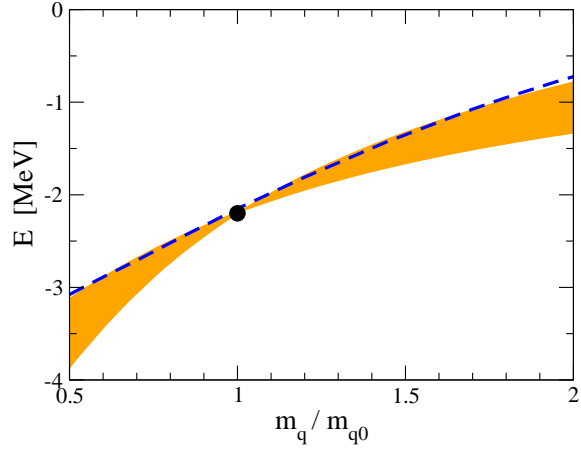}
\end{center}
\caption{Quark-mass dependence of the deuteron binding energy. The
  dashed line corresponds to the LO of the modified Weinberg approach
and the light-shaded band to N$^2$LO result from Ref.~\cite{Berengut:2013nh}.
The band corresponds to the cutoff variation.
The solid dot shows the deuteron binding energy at the physical value
of the quark mass.}
\label{fig:1}       
\end{figure*}

As our new framework is perturbatively renormalizable and we do not
attempt to integrate out the momentum scale $\sim \sqrt{m M_\pi}$, see Ref.~\cite{Mondejar:2006yu},
there is no implicit quark- (pion-) mass dependence of
coupling constants associated with contact interactions. Therefore,
we can straightforwardly calculate the quark-mass dependence of
two-nucleon observables  order-by-order in the chiral expansion.
Below we present an exploratory investigation of chiral extrapolations of the
deuteron binding energy and the scattering lengths in the  $^1S_0$ and
$^3S_1$ partial waves at LO.

\begin{figure}[t]
\includegraphics[width=0.9\textwidth]{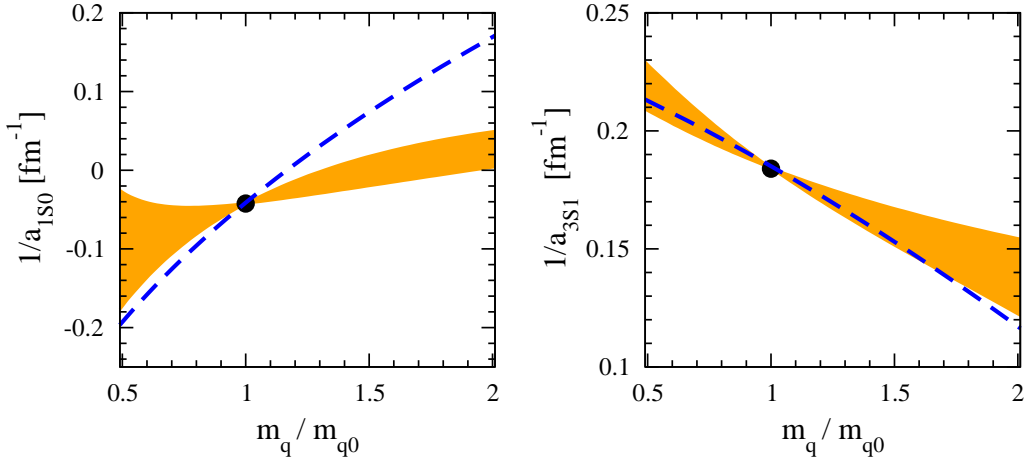}
\caption{Quark-mass dependence of the inverse S-wave scattering
  lengths of $^1S_0$ and $^3S_1$ partial waves. The dashed lines
  correspond
to the LO of the modified Weinberg approach and the light-shaded bands
to N$^2$LO results from Ref.~\cite{Berengut:2013nh}.
The bands correspond to the cutoff variation. The solid dots show the
inverse scattering lengths at the physical value of the quark mass.}
\label{fig:2}       
\end{figure}

There is one free parameter in each of the
S-waves at this order. These parameters are given by linear combinations of the
low-energy constants $C_S$ and $C_T$ in Eq.~(\ref{LOV}) and are
fitted to phase shifts of the Nijmegen partial wave analysis at low
energy, see  Ref.~\cite{Epelbaum:2012ua} for more detail. The quark- or,
equivalently, pion-mass dependence of the NN amplitude at LO is entirely driven
by the explicit pion-mass dependence of the
OPE potential.  Figure \ref{fig:1} shows the resulting  quark-mass
dependence of the deuteron binding energy
together with the recent results of Ref.~\cite{Berengut:2013nh}, see
also Refs.~\cite{Epelbaum:2002gb,Beane:2002vs,Chen:2010yt,Soto:2011tb} for
some earlier EFT calculations along these lines.
Given the theoretical accuracy of our LO analysis and of the
calculation of Ref.~\cite{Berengut:2013nh} which relies on the resonance
saturation hypothesis for contact interactions, the agreement can be
regarded as excellent. We have also calculated the quark-mass
dependence of the inverse scattering lengths which is shown in
Fig. ~\ref{fig:2}.
For the singlet scattering length we observe the same qualitative
behavior as found in Ref.~\cite{Berengut:2013nh}. Somewhat larger
deviations from the N$^2$LO analysis of that work are not surprising
in view of the fact that the OPE potential plays a fairly minor role
in this channel.

\section{Summary}

In this conference contribution we outlined the modified formulation
of baryon chiral perturbation theory for nucleon-nucleon interaction
\cite{Epelbaum:2012ua} and analyzed iterations of the leading order
one-pion exchange potential. Naively, it might appear that the nucleon
mass plays the role of an ultraviolet cutoff in our approach. We have
shown that this interpretation is misleading as it
ignores the fact that the heavy-baryon expansion does not commute with
the expansion in inverse powers of the cutoff parameter.
For one- and two-loop diagrams corresponding to iterations of the OPE
potential, we explicitly demonstrated that renormalization indeed
removes all nucleon-mass dependence which violates the
power counting.

As an application of our approach, we explored the quark-mass dependence
of the deuteron binding energy and the S-wave scattering lengths at LO
in the EFT expansion. The obtained results are in a good agreement
with the recent calculation of Ref.~\cite{Berengut:2013nh}.

\section*{Acknowledgments}
This work is
supported by the EU
(HadronPhysics3 project  ``Study of strongly interacting matter''),
the European Research Council (ERC-2010-StG 259218 NuclearEFT)
the DFG (GE 2218/2-1) and
the Georgian Shota Rustaveli National Science Foundation (grant 11/31).

\end{document}